\documentclass[pre,twocolumn,citesort,aps,floats]{revtex4}%
\usepackage{amsfonts}
\usepackage{amsmath}
\usepackage{amssymb}
\usepackage{graphicx}
\providecommand{\U}[1]{\protect\rule{.1in}{.1in}}
\providecommand{\U}[1]{\protect\rule{.1in}{.1in}}
\providecommand{\U}[1]{\protect\rule{.1in}{.1in}}
\providecommand{\U}[1]{\protect\rule{.1in}{.1in}}

\begin{document}
\title[Steric effects in the dynamics of electrolytes II]
{Steric effects in the dynamics of electrolytes at large applied
voltages:\\II. Modified Poisson-Nernst-Planck equations}
\author{Mustafa Sabri Kilic}
\author{Martin Z. Bazant}
\affiliation{Department of Mathematics, Massachusetts Institute of
Technology, Cambridge, MA 02139,USA.}
\author{Armand Ajdari}
\affiliation{Labortoire de Physico-Chimie Theorique, UMR ESPI-CNRS
7083, 10 rue Vauquelin, F-75005 Paris, France.}

 \pacs{}

\begin{abstract}
  In situations involving large potentials or surface charges, the
  Poisson Boltzman(PB) equation has shortcomings because it neglects
  ion-ion interactions and steric effects. This has been widely
  recognized by the electrochemistry community, leading to the
  development of various alternative models resulting in different
  sets "modified PB equations", which have had at least qualitative
  success in predicting \textit{equilibrium} ion distributions. On the
  other hand, the literature is scarce in terms of descriptions of
  concentration \textit{dynamics} in these regimes.  Here, adapting
  strategies developed to modify the PB equation, we propose a simple
  modification of the widely used Poisson-Nernst-Planck (PNP)
  equations for ionic transport, which at least qualitatively accounts
  for steric effects.  We analyze numerical solutions of these MPNP
  equations on the model problem of the charging of a simple
  electrolyte cell, and compare the outcome to that of the standard
  PNP equations. Finally, we repeat the asymptotic analysis of Bazant,
  Thornton, and Ajdari~\cite{bazant2004} for this new system of
  equations to further document the interest and limits of validity of
  the simpler equivalent electrical circuit models introduced in Part
  I for such problems.
\end{abstract}
\date{\today}
\maketitle

\section{Introduction}

In Part I of this series, we focused on steric effects of finite
ion size on the charging dynamics of a {\it quasi-equilibrium}
electrical double layer, motivated by the breakdown of the
classical Gouy-Chapman model~\cite{lyklema_book_vol2} at large
voltages (up to several Volts and $\gg kT/e = 25$ mV), e.g. which
are commonly applied in AC
electro-osmosis~\cite{ramos1999,brown2001,studer2004,levitan2005,urbanski2006}.
We introduced two simple modifications of the Boltzmann
equilibrium distribution of ions to incorporate steric
constraints. Both new models predicted similar dramatic
consequences of steric effects at large voltages, such as greatly
reduced diffuse-layer capacitance and neutral salt uptake from the
bulk compared to the classical theory. The crucial effect of the
finite ion size is to prevent the unphysical crowding of
point-like ions near the surface at large voltages by forming a
condensed layer of ions at the close-packing limit (likely to
include at least a solvation shell around each ion).

The idea that the electric double layer acts like a capacitor is
part of a bigger picture and suggests that the dynamics can be
described in terms of equivalent circuits
~\cite{macdonald1990,geddes1997}, where the double layer remains
in quasi-equilibrium with the neutral bulk. This classical
approximation has been discussed and validated in the thin double
layer limit by asymptotic analysis of the Poisson-Nernst-Planck
(PNP) equations~\cite{bazant2004}. The PNP equations provide the
standard description of the linear-response dynamics of
electrolytes perturbed from equilibrium, based on the same
assumption of a dilute solution of point-like ions interacting
through a mean field which underlies the PB equation for
equilibrium~\cite{newman_book,lyklema_book_vol2}.

Here we try to account for the effect of steric constraints on the
dynamics, by first establishing modified PNP (MPNP) equations
using linear response theory and modified electrochemical
potentials. We apply the MPNP equations to describe the charging
of a parallel plate electrolyte cell in response to a suddenly
applied voltage and comment on the differences with usual PNP
dynamics. The results are in line with the work in Part I since
the double layer behaves like a capacitor; however its capacitance
is reduced by steric effects, and neutral salt uptake is decreased
as well.

Finally, following the analysis of Bazant, Thornton, and Ajdari
\cite{bazant2004}, we demonstrate that the considerations of Part
I can rigorously be supported by asymptotic analysis on the MPNP
equations. This helps us understand the limits of the electric
double layer capacitor models and define rigorously what is meant
by the thin double layer limit.

\subsection{Electrolyte Dynamics in Dilute Solution Theory}

As we have mentioned in Part I, the dilute solution theory has
been the default model for ion transport for the most part of the
twentieth century. According to this theory, it is acceptable to
neglect interactions between individual ions. As a result, the
electrochemical potential takes the form%
\begin{equation}
\mu_{i}^{dilute}=kT\ln c_{i}+z_{i}e\psi\label{dilchem}%
\end{equation}
where $z_{i}e$ is the charge, $c_{i}$ the concentration, and
$\psi$ the electrostatic potential-usually governed by the
Poisson's equation. In order to derive an equation for the
electrolyte dynamics, we need to combine the above equation with
the flux formula (here with the standard assumption that
interactions between different species are negligible, see e.g.
\cite{bard_book, newman1965, rubinstein2000})
\begin{equation}
\mathbf{F}_{i}=-b_{i}c_{i}\nabla\mu_{i} \label{flux}%
\end{equation}
(where $b_{i}$ is the mobility of the species $i$) and the general
conservation law,%
\begin{equation}
\frac{\partial c_{i}}{\partial t}=-\nabla\cdot F_{i} \label{dynamics}%
\end{equation}
The resulting equations,
\begin{equation}
\frac{\partial c_{i}}{\partial t}=\nabla\cdot\left(  D_{i}\nabla c_{i}%
+b_{i}z_{i}ec_{i}\nabla\psi\right)  , \label{NP}%
\end{equation}
are called the Nernst-Planck (NP) equations. Here, the Einstein's
relation, $D_{i}=b_{i}/kT,$ relates the ions mobility $b_{i}$ to
its diffusivity
$D_{i}.$ The system is closed by the Poisson's equation,%
\begin{equation}
-\nabla\cdot\left(  \varepsilon\nabla\psi\right)  =\sum_{i}z_{i}e
c_{i}
\label{P}%
\end{equation}
where $\varepsilon$ denotes the permittivity of the electrolyte.
The name Poisson-Nernst-Planck(PNP) equations is coined for the
system defined by equations (\ref{NP}) and (\ref{P}). The standard
PNP system presented above has been used to model selectivity and
ionic flux in biological ion channels
\cite{eisenberg1999,kurnikova1999,cardenas2000,hollerbach2000,hollerbach2002},
AC pumping of liquids over electrode
arrays~\cite{ramos1999,ajdari2000,gonzalez2000,brown2001,ramos2003,bazant2006}%
, induced-charge electro-osmotic flows around metallic
colloids~\cite{gamayunov1986,murtsovkin1996} and
microstructures~\cite{iceo2004b,levitan2005,bazant2006},
dielectrophoresis~\cite{shilov1981,simonova2001,squires2006} and
induced-charge electrophoresis
\cite{iceo2004a,yariv2005,squires2006,saintillan2006} of
polarizable particles in electrolytes.

As explained thoroughly in Part I, however, the dilute solution
theory, including the PNP equations, has limited applicability:
its predictions easily violate its basic assumption (i.e. being
dilute) near surfaces of high potential. In fact, this happens
more often than not due to the exponential (Boltzmann-type)
dependence of counter-ion concentration on electrostatic
potential. The steric limit, that is $c_{max}=1/a^{3}$, $a$ being
the typical spacing between densely packed ions, is reached at the critical potential%
\begin{equation}
\Psi_{c}=-\frac{kT}{ze}\ln(c_{0}a^{3})=\frac{kT}{ze}\ln(\frac{c_{\max}}{c_{0}%
}) \label{breakdown}%
\end{equation}
where $c_{0}$ is the bulk electrolyte concentration of either of
the species. Due to the logarithmic dependence in its formulation,
the critical voltage $\Psi_{c}$ is no more than a few times the
thermal voltage $\psi_{T}=kT/ze$ and therefore easily reached in
many applications such as the induced charge electroosmosis
\cite{gamayunov1986,murtsovkin1996}. This has motivated
researchers to modify the standard equations and improve the
dilute solution theory (see next subsection).

\subsection{Beyond Dilute Solution Theory}

Statistical mechanics have proved to be an indispensable tool for
analyzing and improving the dilute solution theory. In particular,
a statistical model with the desired level of detail can be set
up, and after the corresponding free energy $F$ is calculated, the
chemical potentials can be obtained from the formula%
\begin{equation}
\mu_{i}=\frac{\delta F}{\delta c_{i}} \label{free}%
\end{equation}
where $c_{i}$ is the concentration of the $i$-th species. Then
differential equations governing the electrolytes can easily be
derived from this chemical potential as outlined in the previous
subsection. There have been many attempts to calculate the free
energy of the electrolyte more accurately to improve the dilute
solution theory (see below for references). In some of these
attempts, the calculation of the free energy was replaced by an
equivalent consideration of mean electrostatic potential and mean
charge density
\cite{levine1978,outhwaite1980,carnie1981,outhwaite1982,outhwaite1983,carnie1984}%
. Using those mean quantities, various correlation functions as
well as new and more accurate PB (i.e. MPB) equations have been
proposed. However, one should keep in mind that the corresponding
free energies can still be calculated for these models, too.

Perhaps the first examination of the limits of the dilute solution
theory by statistical mechanical considerations was by Kirkwood
\cite{kirkwood1934} in 1934. After a detailed analysis of the
approximations of the dilute solution theory, Kirkwood concluded
that those approximations consisted of the neglect of an
exclusion-volume term and a fluctuation term. Furthermore, he gave
estimates of those terms and argued that they are indeed
negligible in the bulk electrolyte. In recent years, there have
been many attempt
\cite{levine1978,outhwaite1980,carnie1981,outhwaite1982,outhwaite1983,carnie1984}
originating from the liquid-state theory to calculate those
neglected terms more explicitly, which resulted in a variety of
MPB equations (including the MPB1,...,MPB5,\... hierarchy of
Outhwaite and Bhuiyan \cite{outhwaite1980}).

Another general approach to the statistical mechanics of
electrolytes is based on density functional theory (DFT). Using
this formalism,
Rosenfeld~\cite{rosenfeld1989,rosenfeld1994,rosenfeld1997}
systematically derived elaborate free energy functionals for
neutral and charged hard-sphere liquids starting from basic
geometric considerations. Gillespie et al.
\cite{gillespie2002,gillespie2003,gillespie2005} calculated
chemical potentials from Rosenfeld's free energy functionals, and
used them along with the formula (\ref{flux}) to calculate the
steady flux in an ion channel, as well as to investigate the
equilibrium structure of the double layer. With this theoretical
framework, Roth and Gillespie \cite{roth2005} were able to explain
size selectivity of biological ion channels.

Perhaps one reason why neither the MPB hierarchy of Outhwaite and
Bhuiyan \cite{outhwaite1980}, nor the free energy functionals of
Rosenfeld \cite{rosenfeld1997}, has gained widespread use and
recognition is their intrinsic complexity which limits their
simple application to specifc problems. For example, the hard
sphere component of Rosenfeld's free energy density is given by%
\begin{align*}
\Phi_{HS}  &  =-n_{0}\ln\left(  1-n_{3}\right) \\
&  +\frac{n_{1}n_{2}-n_{V1}n_{V2}}{1-n_{3}}+\frac{n_{2}^{3}}{24\pi
(1-n_{3})^{2}}\left(  1-\frac{n_{V2}^{2}}{n_{2}^{2}}\right)  ^{3}%
\end{align*}
where each of $n_{0},n_{1},n_{2},n_{3},n_{V1},$ and $n_{V2}$ are
functions defined in terms of (in general 3D) integrals. So the
improvements over the initial PB equations are made at the cost of
losing the possibility of analytical progress, and nontrivial
numerical work is required even for simple problems in one
dimension.

A considerably simpler approach, which mainly focuses on the
contribution of size effects to the free energy, is based on
mean-field theory together with the lattice-gas approximation in
statistical mechanics. Following the early work of Eigen and Wicke
\cite{wicke1952,freise1952,eigen1954}, Iglic and
Kral-Iglic~\cite{iglic1994,kralj-iglic1996,bohinc2001,bohinc2002}
and Borukhov, Andelman, and
Orland~\cite{borukhov1997,borukhov2000,borukhov2004} were able to
come up with a simple statistical mechanical treatment which
captures basic size effects. A free energy functional is derived
by mean-field approximations of the entropy of equal-sized ions
and solvent molecules. This free energy is then minimized to
obtain the equilibrium average concentration fields, and the
corresponding modified PB (MPB) equation. While more complex than
the original PB equation, it is still rather compact and simple,
and definitely amenable to further analytical use.

\subsection{Scope of the present work}
All of the above authors, as well as others we have cited in Part
I, focus on the equilibrium or steady state properties of the
electrolytes. In fact, we are not aware of any attempt to go
beyond the dilute solution theory (PNP equations) in analyzing the
dynamics of electrolytes in response to time-dependent
perturbations, such as AC voltages. Here in the second part of
this series, our aim is to improve the (time-dependent) PNP
equations of the dilute solution theory by incorporating the
steric effects in a simple way with the goal of identifying new
generic features. Our focus is electrolyte systems that contain
highly charged surfaces, such as an electrode applying a large
voltage $V\gg \psi_{T}=kT/ze$ , where the steric effects become
important quantitatively as well as qualitatively. As in Part I,
here we again adopt the mean-field approach of Borukhov et
al.\cite{borukhov1997} to the size effects, because of two main
reasons: First, and foremost, it is preferable to start with
simple formulations that capture the essential physics while
remaining analytically tractable, as we are mainly interested in
new qualitative phenomena. Second, it is not clear to us how well
the liquid-state theories would perform at large, time-dependent
voltages, since they are (at least in some cases) based on
perturbative methods around an equilibrium reference state -which
may not even exist, say, in presence of an AC electric field.

A quick outline of the paper is as follows: In section II, we
derive the modified Poisson Nernst-Planck (MPNP) equations, using
the free energy obtained by \cite{iglic1994,borukhov1997} to
derive the modified PB equations (MPB). Instead of minimizing the
free energy $F$ we first compute the chemical potentials from the
equation (\ref{free})$.$ Then we describe the electrolyte dynamics
by linear response relations described by equations (\ref{flux}),
(\ref{dynamics}). As a result, we end up with the promised MPNP
equations which include steric corrections to the standard Nernst
Planck equation that become increasingly important as the
concentration field gets large. We continue in section III by
setting up and investigating the numerical solutions of our
modified PNP equations for the problem of parallel plate blocking
electrodes. In section IV, we follow the same lines as in
\cite{bazant2004}, and establish our earlier conclusions
(including the electric circuit picture) about electrical double
layers in Part I by a rigorous asymptotic analysis. We also
calculate higher order corrections to the thin double layer limit,
and check a posteriori the validity of the leading order
approximation. Finally, we close in section V by some comments and
possible future directions for research.

\section{Derivation of Modified PNP Equations}

In this section, we derive the modified PNP equations as outlined
in the introduction. For simplicity, let us restrict ourselves to
the symmetric $z:z$ electrolyte case. We also assume that the
permittivity $\varepsilon$ is constant in the electrolyte. In the
mean-field approximation, the total free energy, $F=U-TS,$ can be
written in terms of the local electrostatic potential $\psi$ and
the ion concentrations $c_{\pm}.$ Following \cite{borukhov1997},
we write the electrostatic energy contribution $U$ as
\begin{equation}
U=\int d\mathbf{r}\left(  -\frac{\varepsilon}{2}\left\vert \nabla
\psi\right\vert ^{2}+zec_{+}\psi-zec_{-}\psi\right)
\end{equation}
The first term is the self-energy of the electric field for a
given potential applied to the boundaries (which acts as a
constraint on the acceptable potential fields), the next two terms
are the electrostatic energies of the ions. The entropic
contribution (the steric effects) can be modeled as
\cite{borukhov1997}%

\begin{eqnarray}
-TS  &  =\frac{kT}{a^{3}}\int d\mathbf{r}\{c_{+}a^{3}\ln\left(  c_{+}%
a^{3}\right)  +c_{-}a^{3}\ln\left(  c_{-}a^{3}\right) \nonumber \\
&  +\left(  1-c_{+}a^{3}-c_{-}a^{3}\right)  \ln\left(  1-c_{+}a^{3}-c_{-}%
a^{3}\right)  \}
\end{eqnarray}
where we have assumed for simplicity that both types of ions and
solvent molecules have the same size $a$. The first two terms are
the entropies of the positive and negative ions, whereas the last
term is the entropy of the solvent molecules. It is this last
term, which penalizes large ionic concentrations.

Requiring that the functional derivatives of this free energy $F$
with respect to $\psi$ and $c_{\pm}$ be respectively $zero$ and
constant chemical potentials $\mu_{\pm}$ (the Lagrange multipliers
for the conserved number of particles of each kind), Borukhov et
al.\cite{borukhov1997} obtain the
modified PB equation%
\begin{equation}
\nabla^{2}\psi=\frac{zec_{0}}{\varepsilon}\frac{2\sinh\left(
\frac{ze\psi }{kT}\right)  }{1+2\nu\sinh^{2}\left(
\frac{ze\psi}{2kT}\right)  }
\label{MPB}%
\end{equation}
Here we go one step further and derive MPNP by calculating the
chemical potentials $\mu_{\pm}$ from (\ref{free}), yielding%
\begin{eqnarray}
\mu_{\pm}  &  =\frac{\delta F}{\delta c_{\pm}}\\
&  =\pm ze\psi+kT(\ln c_{\pm}-\ln\left(
1-c_{+}a^{3}-c_{-}a^{3}\right)  ) \nonumber
\end{eqnarray}
and as a reasonable form for the dynamics we postulate%
\begin{equation}
\frac{\partial c_{\pm}}{\partial\tau}=\nabla\cdot\left(  b_{\pm}c_{\pm}%
\nabla\mu_{\pm}\right)  \label{preMPNP}%
\end{equation}
This form which is classical for close to equilibrium transport is
already an approximation, as it neglects cross-terms in the
mobility matrix (i.e. that the gradients in $\mu_{-}$ can induce a
current of positive ions). Further, we now assume that the
mobilities for each type of ions are the same, and equal to
$b=b_{+}=b_{-}$ which is consistent with the assumption that they
have the same effective size $a$. A final approximation is that we
take this value $b$ to be constant, and in particular insensitive
to the crowding that can occur in the electric double layers. All
these approximations will be further discussed and challenged in
subsequent work, but we proceed for the time being with the
present simpler version, which is a first attempt to incorporate
steric effects in the dynamics. As the electric fields adjusts
almost instantaneously to minimize the electrostatic energy, we
get Poisson equation as
\begin{equation}
\frac{\delta F}{\delta\psi}=\varepsilon\nabla^{2}\psi+ze\left(  c_{+}%
-c_{-}\right)  =0
\end{equation}
The equations (\ref{preMPNP}) yield modified Nernst-Planck
equations:
\begin{align}
\frac{\partial c_{\pm}}{\partial\tau}=  & \\
D\nabla^{2}c_{\pm}\pm &  \frac{D}{k_{B}T}
ze\nabla\cdot(c_{\pm}\nabla
\psi)+a^{3}D\nabla\cdot(\frac{c_{\pm}\nabla\left(
c_{+}+c_{-}\right) }{1-c_{+}a^{3}-c_{-}a^{3}})\nonumber
\end{align}
where $D=kTb$ is the common diffusion coefficient. As mentioned
before, the extra last term is a correction due to the finite size
effects. Considered together the above set of equations are
modified Poisson-Nernst-Planck equations (MPNP), which is our
simplest proposal for a dynamic description that incorporate
steric effects. As in the standard case, this set of equations are
completed by appropriate boundary conditions. In particular, we
consider in this paper that no reactions take place at the surface
(blocking electrodes) so that no-flux boundary conditions
\begin{equation}
D\nabla c_{\pm}\pm
bzec_{\pm}\nabla\psi+\frac{a^{3}Dc_{\pm}\nabla\left(
c_{+}+c_{-}\right)  }{1-c_{+}a^{3}-c_{-}a^{3}}=0
\end{equation}
hold for the ions. We write the boundary condition for the
potential by accounting as before for the possible presence of a
thin insulating layer of fixed capacitance $C_{s}$. This leads to
a mixed boundary condition
\begin{equation}
\psi(n=0)=\psi_{electrode}+\lambda_{S}\frac{\partial\psi}{\partial
n}(n=0)
\label{stern}%
\end{equation}
where $\psi_{electrode}$ is the applied potential at the electrode
which is reduced by the insulating layer to $\psi_{n=0}$ at the
surface of the electrolyte (where MPNP starts to be applied).
$\lambda_{S}=\varepsilon/C_{s}$ is a measure of the thickness of
this layer. Here, $n$ is the normal direction to the surface
pointing into the electrolyte.

\section{Model Problem for the Analysis of the Dynamics}

\begin{figure}
\includegraphics[width=3in]{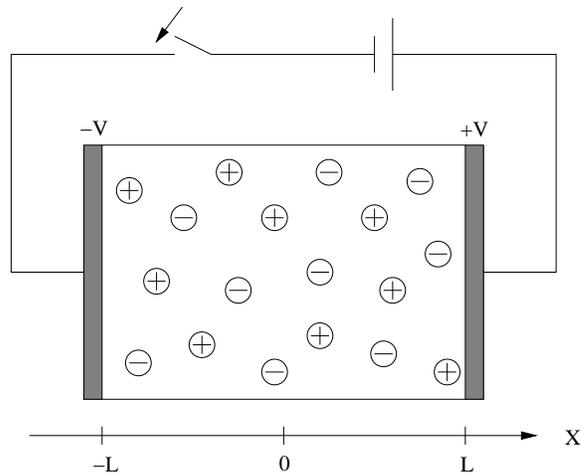}
\caption{ Sketch of the model problem (from
Ref.~\cite{bazant2004}). A
  voltage $2V$ is suddenly applied to a dilute, symmetric, binary
  electrolyte between parallel-plate, blocking electrodes separated by
  $2L$. \label{fig:cartoon} }
\end{figure}

In order to gain some insight into the ramifications of the extra
term we introduced into PNP, we now turn back to the basic model
problem discussed in reference \cite{bazant2004}. Namely, as shown
in Fig.~\ref{fig:cartoon}, we consider the effectively
one-dimensional problem of an electrolyte cell bounded by two
parallel walls (at $x=\pm L$), filled with a z:z electrolyte, at
concentration $c_{0}$, and across which a step voltage of
amplitude ($2V$) is suddenly applied at $t=0$. We further assume
that no Faradaic reactions are induced at the electrodes surface.
We formulate MPNP in this setting, and then compare some numerical
solutions of MPNP to those of PNP. In the next section we will
focus on the case where the double layer
thickness $\lambda_{D}= (2e^{2}%
c_{0}/\varepsilon k_{B}T)^{-1/2}$ is much smaller than the (half)
cell thickness $L$.

First off, we note that, in this simple geometry, the gradients
are replaced by $\frac{\partial}{\partial x}$ and the derivative
with respect to surface normal is replaced by
$\frac{\partial}{\partial x}$ at $x=-L,$ and by
$-\frac{\partial}{\partial x}$ at $x=L.$ Following
\cite{bazant2004}, we cast the MPNP equations in dimensionless
form using $L$ as the reference length scale, and
$\tau_{c}=\lambda_{D}L/D$ as the reference time scale, thus time
and space are represented by $t=\tau D/\lambda_{D}L$ and
$\tilde{x}=x/L.$ The problem is better formulated through the
reduced variables $c=\frac{1}{2c_{0}%
}\left(  c_{+}+c_{-}\right)  $ for the local salt concentration,
$\rho =\frac{1}{2c_{0}}\left(  c_{+}-c_{-}\right)  $ for the local
charge density, and $\phi=ze\psi/kT$ for the electrostatic
potential. The solution is determined by three dimensionless
parameters: $\upsilon=zeV/kT,$ the ratio of the applied voltage to
thermal voltage, $\epsilon=\lambda_{D}/L,$ the ratio of the Debye
length to the system size, $\delta=\lambda_{S}/\lambda_{D}$
introduced in section II which measures the surface capacitance,
and finally a new parameter quantifying the role of steric effects
$\nu=2a^{3}c_{0},$ the effective volume fraction of the ions at no
applied voltage. After dropping the tildes from the variable $x$,
the dimensionless equations take the form
\begin{align*}
\frac{\partial c}{\partial t}  &
=\epsilon\frac{\partial}{\partial x}\left( \frac{\partial
c}{\partial x}+\rho\frac{\partial\phi}{\partial x}+\frac{\nu
c}{1-\nu c}\frac{\partial c}{\partial x}\right) \\
\frac{\partial\rho}{\partial t}  &
=\epsilon\frac{\partial}{\partial x}\left(
\frac{\partial\rho}{\partial x}+c\frac{\partial\phi}{\partial
x}+\frac{\nu\rho}{1-\nu c}\frac{\partial c}{\partial x}\right)
\end{align*}

\begin{equation}
-\epsilon^{2}\frac{\partial^{2}\phi}{\partial x^{2}}=\rho\label{potential}%
\end{equation}
with completely blocking boundary conditions at $x=\pm1,$%

\begin{align}
\frac{\partial c}{\partial x}+\rho\frac{\partial\phi}{\partial
x}+\nu\frac
{c}{1-\nu c}\frac{\partial c}{\partial x}  &  =0\label{nofluxbc}\\
\frac{\partial\rho}{\partial x}+c\frac{\partial\phi}{\partial
x}+\nu\frac {\rho}{1-\nu c}\frac{\partial c}{\partial x}  &
=0\nonumber
\end{align}
in addition to (\ref{stern}), which reads%

\begin{equation}
\phi\pm\delta\epsilon\frac{\partial\phi}{\partial x}=\pm\upsilon\label{42}%
\end{equation}
at $x=\pm1$, where again $\delta$ measures the effective thickness
of the surface insulating layer. Because it is impossible to
satisfy all the boundary conditions when $\epsilon=0$, the limit
of vanishing screening length, $\epsilon\rightarrow0,$ is
singular. The total diffuse charge near the
cathode, scaled by $2zec_{0}L$, is%
\begin{equation}
q(t)=\int_{-1}^{0}\rho(x,t)dx
\end{equation}
The dimensionless Faradaic current, scaled to $2zec_{0}D/L$, is
\begin{equation}
j_{F}=\frac{\partial\rho}{\partial x}+c\frac{\partial\phi}{\partial x}%
+\nu\frac{\rho}{1-\nu c}\frac{\partial c}{\partial x}.
\end{equation}

We have numerically solved these equations for various values of
the dimensionless parameters. As a complete description of this
large parameter space would be very lengthy, we focus on a few
situations for which we provide plots meant to illustrate the
differences brought in by accounting for steric effects. We
therefore also plot the outcome of classical PNP in the same
situations (which correspond to $\nu=0$ in the equations). Of
course a systematic difference is that with the MPNP neither the
concentration $c$ nor the charge density $\rho$ ever overcome the
steric limit $1/\nu$.

\begin{figure}
\begin{center}
(a)\includegraphics[height=2.7985in, width=3.4411in]{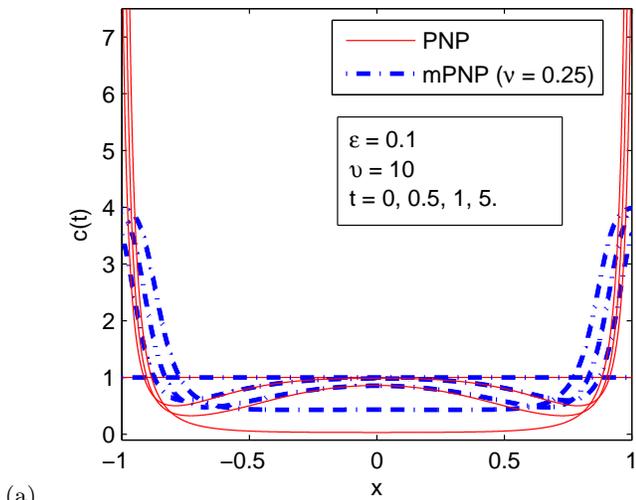} \\
(b)\includegraphics[height=3.55in, width=3.4411in]{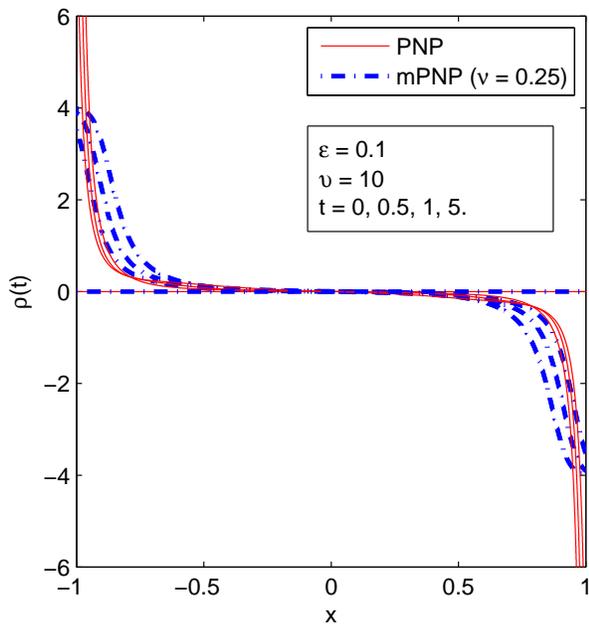}%
\caption{The numerical solutions to the PNP and MPNP systems. The
dimensionless charging voltage is $\upsilon = 10$, which
corresponds to approximately 0.25 Volts at room temperature. The
very large values of $\epsilon=0.1$ and $\nu=0.25$ are chosen
deliberately for illustration. The dimensionless bulk
concentration field $c$ is shown in (a), and the dimensionless
charge density $\rho$ in (b)}%
\label{fig:v10a}%
\end{center}
\end{figure}

For sake of readability of the figure, we start in
Fig.\ref{fig:v10a} with untypically large values for both the
$\epsilon=0.1$ and $\nu=0.25$. The potential is ten times larger
than the thermal voltage $\upsilon=10$. The map of the
concentration and charge density are given for different instants
after the application of the potential drop. Build-up of the
double layers, and the consequent depletion of salt in the bulk
are visible. The MPNP solution stays bounded by $1/\nu$ as
promised, whereas the PNP solution blows up exponentially. A
consequent observation is that salt depletion in the bulk is
weaker with the MPNP, whereas with the classical PNP bulk
concentrations drop to small values even for this moderate
potential ($\sim0.25~V$ in dimensional units). Of course, this is
also a consequence of the large $\epsilon.$

\begin{figure}
\begin{center}
(a)\includegraphics[height=2.7605in, width=3.4411in ]{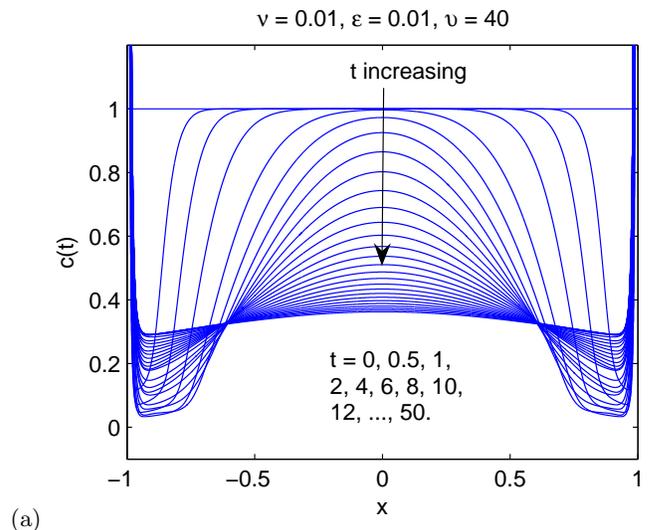}\\
(b)\includegraphics[height=2.5901in, width=3.4411in]{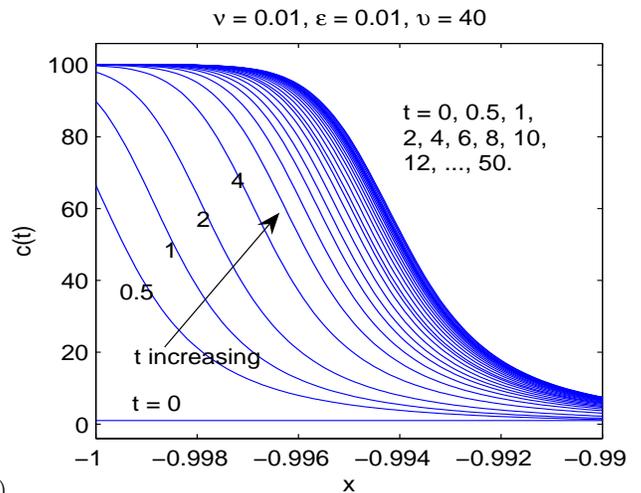}%
\caption{The numerical solution to the MPNP system at the
dimensionless times (i.e. time scaled to the charging time)
$t=0,0.5,1,2,4,6,8,10,...,50.$ The dimensionless charging voltage
is $\upsilon = 40$, which corresponds to approximately 1 Volt at
room temperature. The parameters $\epsilon=\nu=0.01$ are still
large, but comparatively realistic. The dimensionless bulk
concentration field $c$ is shown (a) globally for the
whole region (b) zoomed in at the double layer. }%
\label{fig:v40a}%
\end{center}
\end{figure}

For a simulation with more realistic values, we have taken
$\upsilon=40$ (corresponding to $1~V$ in dimensional units)$,$ and
$\epsilon=\nu=0.01.$ The corresponding solutions are plotted at
nondimensional times $t=0,0.5,1,2,4,6,8,10,...,50$ in
Fig.\ref{fig:v40a}. Figure \ref{fig:v40a}(a) shows the bulk
concentration dynamics, whereas Figure \ref{fig:v40a}(b) focuses
on the double layer near the boundary at $x=-1.$ The MPNP solution
again stays bounded by $1/\nu=100$. The PNP solution is not
plotted as it blows up in the double layer to about
$\cosh(40)\approx10^{17}$ times the bulk value (itself
consequently very small), requiring subtle numerical methods. With
the MPNP, the charge build up of the double layer first proceeds
as it would with the PNP until concentrations close to the
threshold $1/\nu$ are reached. Thereafter, charging proceeds by
growth of the double layer thickness at almost constant density.%

\bigskip

\bigskip

\section{Asymptotic Analysis}


In this section, we adapt to the MPNP equations introduced here
the asymptotic analysis presented in \cite{bazant2004} for PNP
equations in the limit of thin double layers. We show rigourously
that the key properties of the charging dynamics remain the same,
namely, at leading order the boundary layer acts like a capacitor
with a total surface charge density $\tilde{q}(t)$ that changes in
response to the Ohmic current density $\bar{j}(t)$ from the bulk.
However, the expressions for the capacitance of the double layer,
and the diffusive flux from the bulk into the double layer do
change, as we already discussed in Part I.

\subsection{Inner and Outer Expansions}

Adopting the notation in \cite{bazant2004}, we seek regular
asymptotic expansions in $\epsilon$. We observe that the system
has the following symmetries about the origin%
\[
c(-x,t)=c(x,t),\text{ }\rho(-x,t)=-\rho(x,t),\text{ }\phi(-x,t)=-\phi(x,t)
\]
therefore, we consider only $-1<x<0.$

The outer solution, in the bulk, is denoted by a bar accent, and
the expansion takes the form
\[
c=\bar{c}(x,t)=\bar{c}_{0}+\epsilon\bar{c}_{1}+...
\]
It is easy to check that $\bar{c}_{0}=1,$ $\bar{\rho}_{0}=0$ and
$\bar{\phi }_{0}=\bar{j}(t)x$ as a direct consequence of our
choice for the time scale as $\tau_{c}.$ As to the inner solutions
(i.e.$x\approx-1)$, we remove the singularity of equation
$(\ref{potential})$ by introducing $\xi=(1+x)/\epsilon .$ Then, we
obtain
\begin{align}
\epsilon\frac{\partial c}{\partial t} &
=\frac{\partial}{\partial\xi}\left( \frac{\partial
c}{\partial\xi}+\rho\frac{\partial\phi}{\partial\xi}+\frac{\nu
c}{1-\nu c}\frac{\partial c}{\partial\xi}\right)  \label{inner1}\\
\epsilon\frac{\partial\rho}{\partial t} &
=\frac{\partial}{\partial\xi }\left(
\frac{\partial\rho}{\partial\xi}+c\frac{\partial\phi}{\partial\xi
}+\frac{\nu\rho}{1-\nu c}\frac{\partial c}{\partial\xi}\right)  \label{inner2}%
\\
-\frac{\partial^{2}\phi}{\partial\xi^{2}} &  =\rho\label{inner3}%
\end{align}
for which, we can seek regular asymptotic expansions%
\[
c=\tilde{c}(\xi,t)=\tilde{c}_{0}+\epsilon\tilde{c}_{1}+...
\]
Matching inner and outer solutions in space involves the usual van
Dyke conditions, e.g.%
\[
\lim_{\xi\rightarrow\infty}\tilde{c}(\xi,t)=\lim_{x\rightarrow-1}\bar{c}(x,t)
\]
which implies $\tilde{c}_{0}(\infty,t)=\bar{c}_{0}(-1,t),$ $\tilde{c}%
_{1}(\infty,t)=\bar{c}_{1}(-1,t),$ etc. As seen from equations (\ref{inner1}%
)-(\ref{inner3}), there are no terms with time derivatives at
leading order. This quasiequilibrium occurs, because the charging
time $\tau_{c}$ is much larger than the Debye time
$\tau_{D}=\lambda_{D}^{2}/D,$ which is the characteristic time
scale for the local dynamics in the boundary layer.
Consequently, the leading order solution is the "equilibrium"%
\[
\tilde{c}_{0}=\frac{\cosh\tilde{\Phi}_{0}}{1+\nu\left(  \cosh\tilde{\Phi}%
_{0}-1\right)  },\text{ }\tilde{\rho}_{0}=\frac{-\sinh\tilde{\Phi}_{0}}%
{1+\nu(\cosh\tilde{\Phi}_{0}-1)}%
\]
where the excess voltage relative to the bulk
\[
\tilde{\Phi}=\tilde{\phi}(\xi,t)-\bar{\phi}(-1,t)\sim\tilde{\Phi}_{0}%
+\epsilon\tilde{\Phi}_{1}+...
\]
satisfies the modified PB equation at leading order%

\begin{equation}
\frac{\partial^{2}\tilde{\Phi}_{0}}{\partial\xi^{2}}=\frac{\sinh\tilde{\Phi
}_{0}}{1+\nu\left(  \cosh\tilde{\Phi}_{0}-1\right)  } \label{leadpois}%
\end{equation}
with $\tilde{\Phi}_{0}(\infty,t)=0,$ and
$\tilde{\Phi}_{0}(0,t)=\tilde{\zeta }(t),$ the dimensionless zeta
potential, which varies as the diffuse layer charges. Applying
matching to the electric field, we obtain
\begin{equation}
\frac{\partial\tilde{\phi}}{\partial\xi}\left(  \infty,t\right)
\sim \epsilon\frac{\partial\bar{\phi}}{\partial x}(-1,t)\text{
}\Rightarrow\text{
}\frac{\partial\tilde{\phi}_{0}}{\partial\xi}(\infty,t)=0 \label{65}%
\end{equation}
and therefore an integration of (\ref{leadpois}) yields%
\begin{equation}
\frac{\partial\tilde{\Phi}_{0}}{\partial\xi}=-\mbox{sign}(\tilde{\Phi}%
_{0})\sqrt{\frac{2}{\nu}\ln\left(  1+\nu\left(  \cosh\tilde{\Phi}%
_{0}-1\right)  \right)  } \label{66}%
\end{equation}

\subsection{Time-dependent matching}
So far we have found that both the bulk and the boundary layers
are in quasiequilibrium, which apparently contradicts the dynamic
nature of the charging process. This is reconciled by noting once
more that the boundary condition for the inner solution involves
the quantity $\zeta(t),$ which varies in response to the diffusive
flux from the bulk. Motivated by the physics, we consider the
total diffuse charge, which has the scaling
$q(t)\sim\epsilon\tilde{q}(t),$ where%
\begin{equation}
\tilde{q}(t)=\int_{0}^{\infty}\tilde{\rho}(\xi,t)d\xi\sim\tilde{q}%
_{0}+\epsilon\tilde{q}_{1}+\epsilon^{2}\tilde{q}_{2}+...\label{71}%
\end{equation}
Taking a time derivative, and using (\ref{inner2}) together with
the no flux boundary condition (\ref{nofluxbc}), we obtain%
\begin{align*}
\frac{d\tilde{q}}{dt}(t)  &
=\lim_{\xi\rightarrow\infty}\frac{1}{\epsilon }\left(
\frac{\partial\tilde{\rho}}{\partial\xi}+\tilde{c}\frac{\partial\phi
}{\partial\xi}+\frac{\nu\tilde{\rho}}{1-\nu\tilde{c}}\frac{\partial\tilde{c}%
}{\partial\xi}\right) \\
&  \sim\lim_{x\rightarrow-1}\left(  \frac{\partial\bar{\rho}}{\partial x}%
+\bar{c}\frac{\partial\phi}{\partial x}+\frac{\nu\bar{\rho}}{1-\nu\bar{c}%
}\frac{\partial\bar{c}}{\partial x}\right)
\end{align*}
where we applied matching to the flux densities. Substituting the
regular expansions of inner and outer solutions yield a hiararchy
of matching conditions. At leading order, we have%
\begin{equation}
\frac{d\tilde{q}_{0}}{dt}(t)=\bar{j}_{0}(t) \label{73}%
\end{equation}
which, being a balance of $O(1)$ quantities, is reassuring us that
we have chosen the correct time-scale. This is the key equation
which tells us that at leading order, the double layer behaves
like a capacitor, whose total surface charge density $\tilde{q},$
changes in response to the transient Faradaic current density,
$\bar{j}(t),$ from the bulk.

\subsection{Leading Order Dynamics}
Using equations (\ref{inner3}),(\ref{65}) and (\ref{66}), the
integral in (\ref{71}) can be performed at leading order to yield%
\begin{equation}
\tilde{q}_{0}(t)=-\mbox{sign}(\tilde{\zeta}_{0})\sqrt{\frac{2}{\nu}\ln
[1+\nu(\cosh\tilde{\zeta}_{0}-1)]} \label{q0}%
\end{equation}
\bigskip Then the Stern boundary condition, eq.(\ref{42}), yields
\begin{align}
&
\tilde{\zeta}_{0}+\delta\mbox{sign}(\tilde{\zeta}_{0})\sqrt{\frac{2}{\nu
}\ln[1+\nu(\cosh\tilde{\zeta}_{0}-1)]}\nonumber\\
&  =\bar{j}_{0}(t)-\upsilon=\tilde{\Psi}_{0} \label{sternbc}%
\end{align}
where $\tilde{\Psi}(t)=-\upsilon-\bar{\phi}(-1,t)\sim\tilde{\Psi}_{0}%
+\epsilon\tilde{\Psi}_{1}+...$ is the total voltage across the
compact and the diffuse layers. Equation (\ref{sternbc}) results
in higher $\tilde{\zeta }_{0}$ than its classical PNP (or PB)
counterpart
\begin{equation}
\tilde{\zeta}_{0}+2\delta\sinh(\tilde{\zeta}_{0}/2)=\tilde{\Psi}_{0}
\label{classsternbc}%
\end{equation}
because the left hand side of (\ref{sternbc}) is always smaller
than the left hand side of (\ref{classsternbc}) for any given
$\tilde{\zeta}_{0}.$ In both formulas, the first term (i.e.
$\tilde{\zeta}_{0}$) is the voltage drop over the diffuse layer
whereas the second term (i.e. $-\delta\tilde{q}_{0})$ is the
voltage drop over the Stern layer. For high applied voltages, the
standard model assigns exponentially bigger proportions of that
voltage to the Stern layer, whereas the modified theory predicts a
balanced distribution. For more details, see Part I section V.

Substituting into the matching condition (\ref{73}), we obtain an
ordinary initial value problem
\begin{equation}
-\tilde{C}_{0}(\tilde{\Psi}_{0})\frac{d\tilde{\Psi}_{0}}{dt}=\tilde{\Psi}%
_{0}+\upsilon,\text{ }\bar{j}_{0}(0)=\upsilon\label{weaklynonlinear}%
\end{equation}
where
\begin{align}
\tilde{C}_{0}  &  =-\frac{d\tilde{q}_{0}}{d\tilde{\Psi}_{0}}\label{78}\\
&  =\frac{1}{\frac{1+\nu(\cosh\tilde{\zeta}_{0}-1)}{|\sinh\tilde{\zeta}_{0}%
|}\sqrt{\frac{2}{\nu}\ln(1+\nu(\cosh\tilde{\zeta}_{0}-1))}+\delta}\nonumber
\end{align}
is the differential capacitance for the double layer as a function
of its voltage. This has a completely different behaviour than its
counterpart in
\cite{bazant2004}, namely%
\begin{equation}
\tilde{C}_{0}=\frac{1}{\mbox{sech}(\tilde{\zeta}_{0}/2)+\delta} \label{old}%
\end{equation}
especially at higher $\tilde{\zeta}_{0}.$ As
$\tilde{\zeta}_{0}\rightarrow \infty,$ our differential
capacitance
$\tilde{C}_{0}^{\nu}\sim(\sqrt{\frac{2}{\nu}\tilde{\zeta}_{0}}+\delta)^{-1}\rightarrow0,$
in contrast to $\tilde{C}_{0}^{\nu=0},$ the differential
capacitance given by (\ref{old}), which tends to $\delta^{-1}$(see
Part I for more details). Thus, at high zeta potentials, steric
effects decrease the capacitance, possibly down to zero.
Physically, this is because of the increasing double layer
thickness as a result of excessive pile up of the ions coming from
the bulk into the double layer (see Part 1).

Equation (\ref{78}) is separable, and its solution can be expressed in the form%
\begin{equation}
\tilde{\Psi}_{0}=\bar{j}_{0}(t)-\upsilon=-F^{-1}(t) \label{dyn}%
\end{equation}

\begin{equation}
F(z)=\int_{0}^{z}\frac{\tilde{C}_{0}(u)}{\upsilon+u}du \label{func}%
\end{equation}
Steric effects reduce the capacitance, and therefore $F(z)$, which
by formula (\ref{func}) implies a faster relaxation process for
the voltage difference $\tilde{\Psi}_{0}$. In other words, the
$RC$ relaxation time is shortened (see Part I).

\subsection{Neutral Salt Adsorption by the Double Layer}

A natural consequence of reduced capacitance at the electrodes is
the reduction of the amplitude of the diffusion layer in the bulk
were the adsorbed salt in the double layer is extracted from. We
now revisit some of the ideas in \cite{bazant2004}, and
recalculate the neutral salt adsorption by the double layer. The
excess ion concentration, $c-c_{0},$ acquired by the double layers
is accounted for by the diffusion from the bulk at $O(\epsilon)$
or higher, as diffusion is absent at leading order. Following
\cite{bazant2004}, we introduce the variable
$w(t)=\epsilon\tilde{w}(t),$ akin to $q(t),$ which represents the
excess amount of salt in the double layer:
\[
\tilde{w}(t)=\int_{0}^{\infty}\left[  \tilde{c}(\xi,t)-\bar{c}_{0}%
(-1,t)\right]  d\xi=\tilde{w}_{0}(t)+\epsilon\tilde{w}_{1}(t)+...
\]
Taking a time derivative, we find
\begin{align}
\frac{d\tilde{w}_{0}}{dt}(t) &
=\lim_{\xi\rightarrow\infty}\frac{1}{\epsilon}\left(\frac{\partial\tilde{c}}{\partial\xi}+\tilde{\rho}
\frac{\partial\phi}{\partial\xi}+\frac{\nu\tilde{c}}{1-\nu\tilde{c}}\frac{\partial\tilde{c}%
}{\partial\xi}\right)  \nonumber\\
&  \sim\lim_{x\rightarrow-1}\left(  \frac{\partial\bar{c}}{\partial x}%
+\bar{\rho}\frac{\partial\phi}{\partial x}+\frac{\nu\bar{c}}{1-\nu\bar{c}%
}\frac{\partial\bar{c}}{\partial x}\right)  \label{dwdt}%
\end{align}
\bigskip Since $\bar{c}=1+\epsilon\bar{c}_{1}+...$ , at leading order
$\bar{c}\sim1$  and $\frac{\partial\bar{c}}{\partial}\sim\epsilon
\frac{\partial\bar{c}_{1}}{\partial x},$ therefore (\ref{dwdt})
yields
\begin{equation}
\frac{d\tilde{w}_{0}}{d\bar{t}}(t)=\frac{\left(  1-\nu\right)
}{\epsilon}\frac{d\tilde{w}_{0}}{dt}(t)=
\frac{\partial\bar{c}_{1}}{\partial x}(-1,t)\label{95}%
\end{equation}
which involves the new time variable%
\[
\bar{t}=\frac{\epsilon}{\left(  1-\nu\right)
}t=\frac{\epsilon}{\left(
1-\nu\right)  }\frac{\tau}{\tau_{c}}=\frac{\tau}{\tau_{s}}%
\]
scaled to bulk diffusion time, $\tau_{s}=\left(  1-\nu\right)
L^{2}/D,$ which is slightly different from the time scale given in
\cite{bazant2004}. The salt uptake $\tilde{w}(t)$ can be expressed
in terms of an integral
\begin{align}
\tilde{w}(t) &  =\int_{0}^{\infty}\left(  \frac{\cosh\tilde{\Phi}}%
{1+\nu\left(  \cosh\tilde{\Phi}-1\right)  }-1\right)  d\xi\nonumber\\
&  =\int_{0}^{\zeta}\frac{\cosh\tilde{\Phi}-1}{1+\nu\left(
\cosh\tilde{\Phi}-1\right)  }\frac{\left(  1-\nu\right)  d\tilde{\Phi}}{\sqrt{\frac{2}{\nu}%
\ln\left(  1+\nu\left(  \cosh\tilde{\Phi}-1\right)  \right)  }}\label{w}%
\end{align}
with no obvious further simplification.

We now proceed to calculate the depletion of the bulk
concentration during the double-layer charging. Note that the new
time scale $\tau_{s}$ introduced by
(\ref{95}) is the time scale for the first order diffusive dynamics in bulk%
\begin{equation}
\frac{\partial\bar{c}_{1}}{\partial\bar{t}}=
\frac{\left(1-\nu\right )}{\epsilon}
\frac{\partial\bar{c}_{1}}{\partial t}=\frac{\partial^{2}\bar{c}_{1}}{\partial x^{2}}%
\end{equation}

As the source is defined by (\ref{95}) in terms of gradients, an
appropriate Green's function can be obtained by taking Laplace
transforms and using method of images as in \cite{bazant2004}, which leads to%
\begin{equation}
\bar{c}_{1}(x,t)=-\int_{0}^{\bar{t}}d\bar{t}^{\prime}G(x,\bar{t}-\bar
{t}^{\prime})\frac{\partial}{\partial\bar{t}}[\tilde{w}_{0}(\bar{t}^{\prime
}/\epsilon)] \label{solution}%
\end{equation}
where
\begin{equation}
G(x,\bar{t})=
\frac{1}{\sqrt{\pi\bar{t}}}\sum_{m=-\infty}^{\infty}e^{-\left(x-2m+1\right)  ^{2}/4\bar{t}} \label{gaussian}%
\end{equation}

In the limit $\epsilon\rightarrow0,$ the initial charging process
at the time scale $\tau_{c}=O\left(  \epsilon\right)  $ is almost
instantaneous, which is followed by the slow relaxation of the
bulk diffusion layers. This limit corresponds to approximating the
source terms in the integral in (\ref{solution}) to exist only in
a small $O\left(  \epsilon\right)$ neighborhood of zero, or more explicitly%

\begin{align}
\lim_{\epsilon\rightarrow0}\bar{c}_{1}(x,\bar{t}) &  \simeq-G(x,\bar{t}%
)\int_{0}^{\bar{t}}d\bar{t}^{\prime}\frac{\partial}{\partial\bar{t}}[\tilde
{w}_{0}\left(  \frac{\bar{t}^{\prime}}{\epsilon}\right)  ]\nonumber\\
&  =-\tilde{w}_{0}\left(  \infty\right)  G(x,\bar{t})\label{approx}%
\end{align}
with $\tilde{w}_{0}\left(  \infty\right)  =$%
\begin{equation}
\int_{0}^{f^{-1}(\upsilon)}\frac{\cosh\tilde{\Phi}-1}{1+\nu\left(
\cosh \tilde{\Phi}-1\right)}
\frac{\left(1-\nu\right)d\tilde{\Phi}}{\sqrt{\frac{2}{\nu}\ln\left(  1+\nu\left(  \cosh\tilde{\Phi}-1\right)  \right)  }%
}\label{w00}%
\end{equation}
where%
\begin{equation}
f(\zeta)=\zeta+\mbox{sign}(\zeta)\delta\sqrt{\frac{2}{\nu}\ln\left(
1+\nu\left(  \cosh\zeta-1\right)  \right)  }.\label{f}%
\end{equation}
As expected from the underlying physics, and already explained in
Part I and illustrated in the plots of section III above, the
formula (\ref{w00}) predicts smaller values for the the depth of
the bulk diffusion than its classical PNP ($\nu=0)$ counterpart.
This difference is more pronounced at higher $\upsilon.$ Equation
(\ref{approx}) is a simple approximation that describes two
diffusion layers created at the electodes slowly invading the
entire cell. At first,they have simple Gaussian profiles%
\begin{equation}
\bar{c}(x,t)\sim1-
\frac{\epsilon\tilde{w}_{0}\left(\infty\right)}{\sqrt{\pi\bar{t}}}
\left[e^{-\left(x+1\right)^{2}/4\bar{t}}+e^{-\left(
x-1\right)  ^{2}/4\bar{t}}\right]  \label{gausapprox}%
\end{equation}
for $\epsilon<<\bar{t}<<1.$ The two diffusion layers eventually
collide, and the concentration slowly approaches a reduced constant value,%
\begin{equation}
\bar{c}(x,t)\sim1-\epsilon\tilde{w}_{0}\left(  \infty\right)  \label{steady}%
\end{equation}
for $\bar{t}>>1,$ as we expect from the steady-state excess
concentration from the double layers.

Of course one expects that at large enough applied voltage, the
above approximation breaks down, as the decrease in bulk
concentration becomes significant and thus modifies the value of
$w$ in the double layer.

\subsection{Validity of the Weakly Nonlinear Approximation}

The first order solution consisting of the variables indexed by
"0" is often referred to as the \textit{weakly nonlinear
approximation}, whose main feature is that the bulk concentrations
are constant, namely $\bar{c}\approx\bar {c}_{0}=1$ and
$\bar{\rho}\approx\bar{\rho}_{0}=0.$ The system therefore is
characterized only by the surface charge
$\tilde{q}\approx\tilde{q}_{0},$ or the double layer potential
difference $\tilde{\Psi}\approx\tilde{\Psi}_{0},$ which is
governed by the nonlinear ODE in equation (\ref{weaklynonlinear}).
This corresponds to modeling the problem by an equivalent circuit
model with variable capacitance for the double layer, and constant
bulk electrolyte resistance.

In order to understand when the weakly nonlinear approximation
holds, we can compare the size of the next order approximation to
the leading term. Although not a rigorous proof, one may argue
that if $|\epsilon c_{1}|$ is much smaller than $\bar{c}_{0}=1,$
then leading order term is a good approximation to the full
solution. We will get help from the approximations
(\ref{gausapprox}) and (\ref{steady}) to see if this is the case.

Seen in the light of equation (\ref{steady}), the assumption that
the first correction is much smaller than the leading term
requires that $\nu _{b}=\epsilon\tilde{w}_{0}\left(  \infty\right)
\ll1,$ in other words $\epsilon\left(  1-\nu\right)
\int_{0}^{f^{-1}(\upsilon)}\frac{\cosh\Phi-1}{1+\nu\left(  \cosh\Phi-1\right)  }
\frac{d\Phi}{\sqrt{\frac{2}{\nu}%
\ln\left(  1+\nu\left(  \cosh\Phi-1\right)  \right)  }}\ll1,$
where $f$ is given in (\ref{f}). After a series of approximations,
including $\left(1-\nu\right)  \approx1,$ $\delta=O(1),$ and
$\zeta\gg1$(or $\upsilon\gg1),$ this becomes
$\epsilon\sqrt{\frac{2}{\nu}\ln\left(  1+\nu\cosh\upsilon\right)
}\ll1.$ If $\nu$ is not too close to zero(i.e.
$\nu\cosh\upsilon\gg1)$, this is the same as
$2\epsilon^{2}\upsilon/\nu\ll1.$ Putting the units back, we
obtain%
\begin{equation}
2\frac{\lambda_{D}^{2}}{L^{2}a^{3}c_{0}}\frac{zeV}{kT} \ll1 \label{sc}%
\end{equation}
For a typical experiment with $\lambda_{D}=10$ nm$,$ $L=0.1mm,$
$c_{0}=1mM,$ $a=5\mathring{A},$ and at room temperature, this
condition becomes $V \ll188$ Volts.

However, the weakly nonlinear \textit{dynamics }breaks down at
somewhat smaller voltages, because the neutral salt adsorption
causes a temporary, local depletion of bulk concentration
exceeding that of the final steady state. In our model problem,
the maximum change in bulk concentration occurs just outside the
diffuse layers at $x=\pm1,$ just after the initial charging
process finishes at the same scale $t=O(1)$ or
$\bar{t}=O(\epsilon).$ Letting $\bar{t}=\epsilon,$ and $x=\pm1$ in
equation (\ref{gausapprox}), we obtain the first two terms in the
asymptotic expansion as
\[
c\left(  \pm1,\epsilon\right)  =1-\sqrt{\frac{\epsilon}{\pi}}\tilde{w}%
_{0}\left(  \infty\right)
\]
At that time, the double layers have almost been fully charged ,
however the bulk diffusion has only had time to reach a region of
length $O\left(\sqrt{\epsilon}\right)  .$So the concentration is
depleted locally by $O\left(
\frac{\epsilon}{\sqrt{\epsilon}}\right)  =O\left(  \sqrt{\epsilon
}\right)  ,$ which is much more than the uniform $O(\epsilon)$
depletion remaining after complete bulk diffusion.

Therefore, in order for the time-dependent correction term to be
uniformly smaller than the leading term, we need
\begin{equation}
\sqrt{\frac{\epsilon}{\pi}}\tilde{w}_{0}\left(  \infty\right)  \ll1 \label{cc}%
\end{equation}
By the same approximations as in (\ref{sc}), this yields
$\frac{2\epsilon \upsilon}{\sqrt{\pi}\nu} \ll1,$ or with the units
\begin{equation}
\frac{2}{\sqrt{\pi}}\frac{\lambda_{D}}{La^{3}c_{0}} \frac{zeV}{kT}
\ll1
\label{sc2}%
\end{equation}
The corresponding threshold voltage is smaller than the former by
a factor of roughly $\sim L/\lambda_{D}=\epsilon^{-1}$. For the
same set of parameter as above $\lambda_{D}=10$ nm$,$ $L=0.1mm,$
$c_{0}=1mM,$ $a=5\mathring{A},$ and at room temperature, this
condition gives $V \ll0.033$ Volts. Thus according to this
criterion, the weakly nonlinear approximation easily breaks down,
and one may need to consult to the full MPNP system for an
understanding of the electrolyte dynamics.

\begin{figure}
[ptb]
\begin{center}
\includegraphics[height=2.9136in, width=3.4411in]{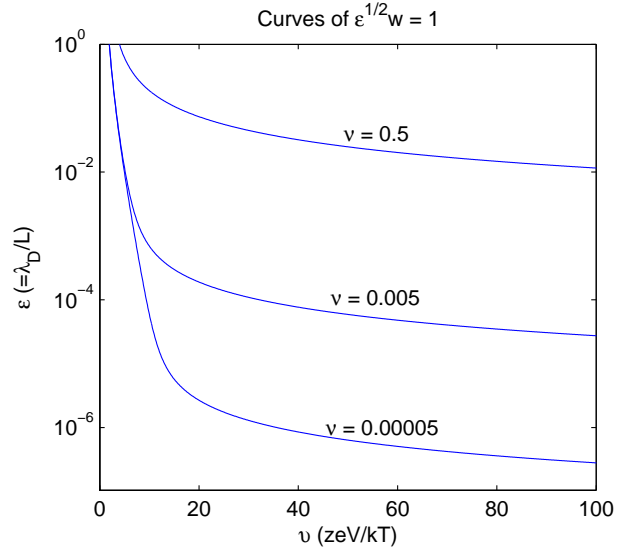} \\
\caption{Shown are the curves on which $\sqrt{\epsilon}w=1$ for
various values of $\nu.$ The weakly nonlinear approximation holds
to the south-west of these
curves when $\sqrt{\epsilon}w<<1.$}%
\label{fig:epsw2}%
\end{center}
\end{figure}

A more accurate understanding into the condition (\ref{cc}) is
gained by the numerical study of the function $w.$ The curves on
which $\sqrt{\epsilon}w=1$ (i.e. $\epsilon=w^{-2}$) are plotted in
Fig.\ref{fig:epsw2} for various values of $\nu$ (here we dropped
the somewhat arbitrary factor $\sqrt{\pi}$ in (\ref{cc})).The
weakly nonlinear approximation holds to the south-west of these
curves when $\sqrt{\epsilon}w\ll1.$ The criterion given by the
inequality (\ref{sc2}) corresponds to the asymptotic behaviour of
those curves as $\upsilon$ tends to infinity.

\begin{figure}
\begin{center}
(I)\includegraphics[height=2.9568in, width=3.44in]{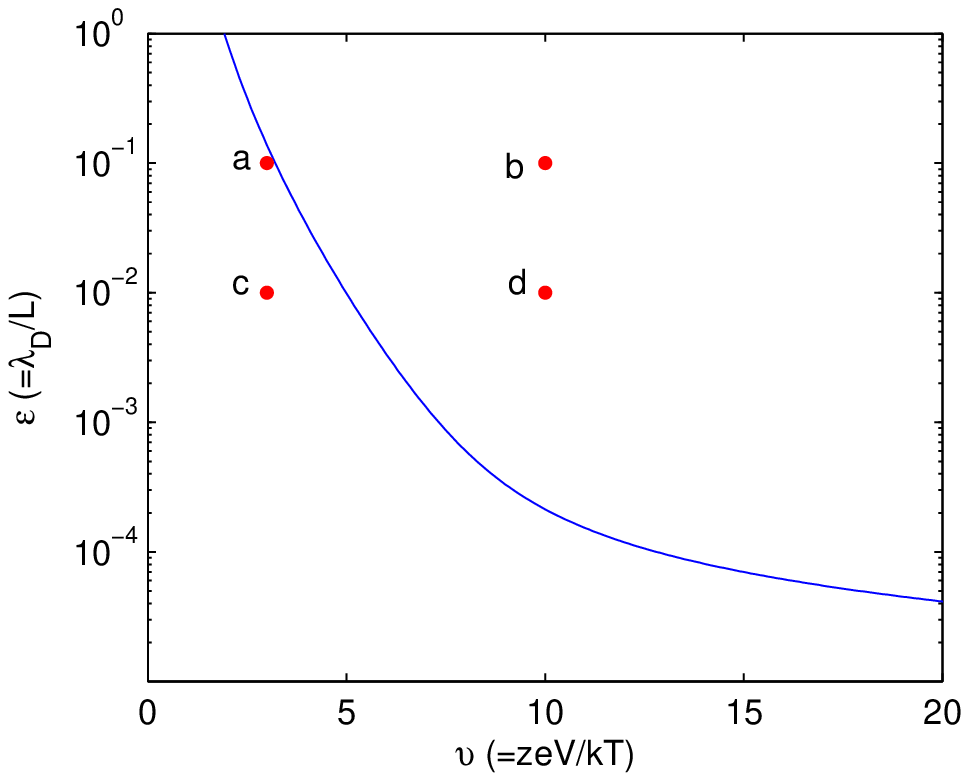} \\
(II)\includegraphics[height=2.9568in, width=3.44in]{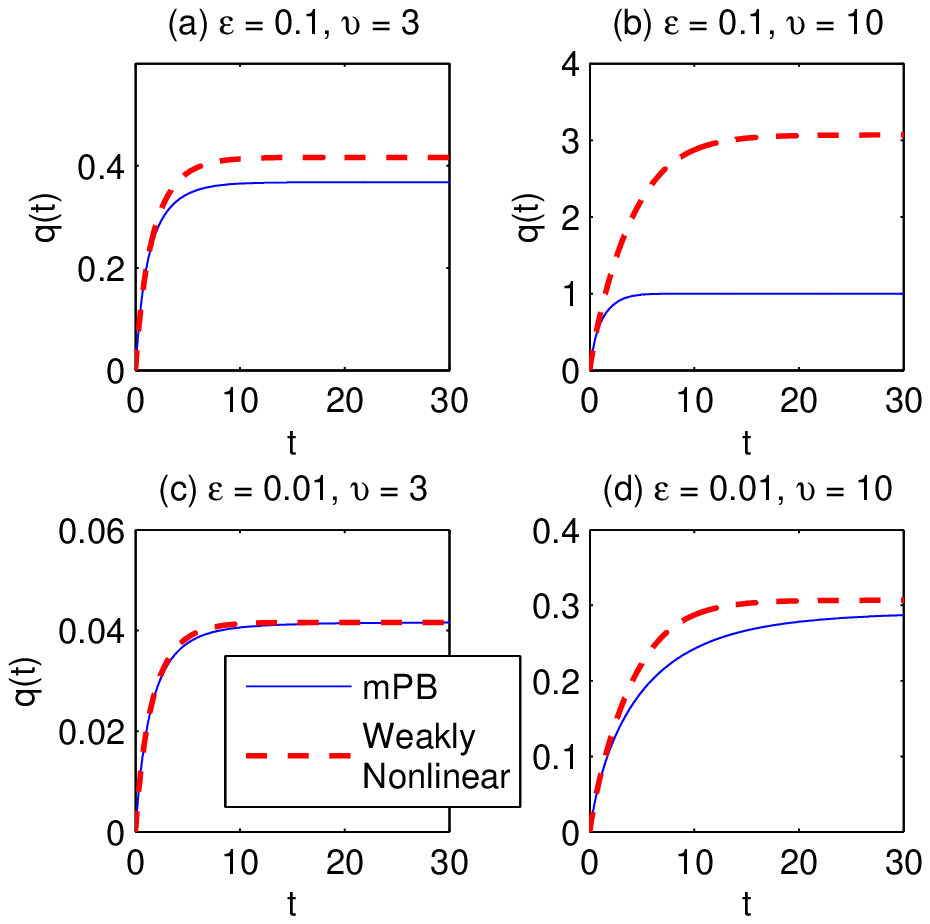}
\caption{Four case studies on the validity of the weakly nonlinear
approximation. No stern capacitance is included ($\delta=0$). (I)
The $\sqrt{\epsilon}w=1$ curve for the particular value
$\nu=0.01,$ and corresponding locations of the four case studies
shown. (II) The comparison of the charge $q(t)$ stored in the half
cell computed by (i) the MPNP (ii) the weakly nonlinear
approximation corresponding to the MPNP. The match in (c) is good
as expected, and the match in (b) is off by several factors, again
as expected. In cases (a) and (d), the curves run close but they
are clearly separated. }%
\label{fig:Qcomp}%
\end{center}
\end{figure}

To observe that this is the case numerically, we have also
compared the charging dynamics given by the weakly nonlinear
approximation to that of the full MPNP solution in
Fig.\ref{fig:Qcomp}. with the parameters $\epsilon =\nu=0.01.$
When the parameters were to the south-west of the $\sqrt{\epsilon
}w=1,$ as in case (c), the match was perfect. In the other cases,
the weakly nonlinear approximation did not do as well, it was
particularly off for case (b), when the product $\sqrt{\epsilon}w$
yielded the highest number. In case (a), the agreement was off by
a constant shift, whereas in case (d), the situation improved over
time. This is because $\epsilon w$ is still small for case (d),
although $\sqrt{\epsilon}w$ is not, and therefore weakly nonlinear
approximation is still valid for the final state of the system. To
summarize, if an accurate description of the system at all times
is desired, then $\sqrt{\epsilon}w<<1$ is the appropriate
criterion, however it may suffice to have just $\epsilon w<<1$ to
be able to predict the eventual steady state by the weakly nonlinear model. %

\bigskip

\section{Conclusions}
As an extension of the modified PB approach, we have derived
modified PNP equations, which may be of help when the thin double
layer approximation fails. Using this new set of equations, we
have revisited the model problem of the step charging of a cell
with parallel blocking electrodes. In addition, we have confirmed
through asymptotic analysis the hypotheses stated in Part I
regarding the MPB double layer model. We have also investigated
the limits of the thin double layer approximation as well as
higher order corrections.

The MPNP system proposed is a natural extension of one of the two
models introduced in Part I, namely the MPB model based on an
approach originally due to \cite{iglic1994,borukhov1997}. One can
similarly construct other MPNP equations from other MPB models
such as the composite diffuse-layer model also introduced in Part
I. However the discontinuous structure of the latter leads to a
complex formalism with discontinuities, improper for
implementation in complex geometries, whereas the one presented
here has a smooth behavior and is therefore much more broadly
applicable.

We expect that the MPNP equations presented here, or other simple
variations with different modifications of the chemical potentials
or free energy, will find many applications. They are no more
difficult to use than the classical PNP equations, which are
currently ubiquitous in the modeling of electrochemical systems.
Especially at large voltages, the MPNP equations are much better
suited for numerical computations, since they lack the
exponentially diverging concentrations predicted by the PNP
equations, which are difficult to resolve. Of course, those same
divergences are also clearly unphysical, while the MPNP equations
predict reasonable ion profile for any applied voltage.

Future research directions include the application of the
presented framework to many other settings including the response
of the simple electrolytic cell considered here to various systems
with time-dependent applied voltages of strong amplitudes. In
particular, driving an electrochemical cell or AC electro-osmotic
pump at rather large frequencies should be a selective way of
checking the validity/use of equations such as the one put forward
here because dynamical effects are exacerbated in such situations.

\newcommand{\noopsort}[1]{} \newcommand{\printfirst}[2]{#1}
  \newcommand{\singleletter}[1]{#1} \newcommand{\switchargs}[2]{#2#1}

\end{document}